# New Light on Dark Matter

Jeremiah P. Ostriker[1,2] and Paul Steinhardt[2]

**ABSTRACT**
**Dark matter, proposed decades ago as a speculative component of the universe, is now known to be the vital ingredient in the cosmos, eight times more abundant than ordinary matter, one quarter of the total energy density and the component which has controlled the growth of structure in the universe. Its nature remains a mystery, but, assuming it is comprised of weakly interacting sub-atomic particles, is consistent with large scale cosmic structure. However, recent analyses of structure on galactic and sub-galactic scales have suggested discrepancies and stimulated numerous alternative proposals. We discuss how studies of the density, demography, history and environment of smaller scale structures may distinguish among these possibilities and shed new light on the nature of dark matter.**

The last decade, has seen the emergence of a standard model, sometimes called a "concordance model" (*1*) to emphasize that its predictions are in concord with current observations – both of the nearby universe and of the early universe as seen in the cosmic background radiation ("CBR"). The new CBR results (*2*) recently published by the team of scientists analyzing the observations of the WMAP satellite are generally assessed as providing a brilliant and comprehensive verification of the concordance model. But consistency or concordance is not scientific proof. The model works. And, many previously proposed models are now known to be wrong in essential elements. Most of the current work in cosmology is focused on pinning down the adjustable parameters in the concordance model to see how precisely we can specify that model and whether any discrepancies will appear as this process is advanced. History, however, is a stern teacher. We know that all physical models for natural phenomena represent approximations to the truth. The blinding light that seems – for the moment – to be illuminating our path is certain to expose stubborn facts that will cause us to modify and extend the current paradigm.

First, according to the standard model, there is ordinary matter consisting of the familiar chemical elements. Nuclear cooking that occurred in the first few minutes after the big bang left a soup containing primarily hydrogen and helium and other light elements. Recent measurements of chemical abundances are consistent with theoretical predictions (*3*) – provided the mass density of ordinary matter is about 4% of the total energy density of the universe and from WMAP we now know that this estimate is accurate. But, we also know from various gravitational effects that the total mass density is much more than 4% of the total energy density.

Over 65 years ago the Swiss astrophysicist Fritz Zwicky (*4*) noticed that the speed of galaxies in large clusters, such as the Coma cluster, is much too great to keep them gravitationally bound together unless they weigh over one hundred times more than one would estimate based on the number of stars in the cluster. Decades of investigation confirmed his analysis, and in the 1970s further evidence for dark matter was found from gravitational studies of matter in the outer parts (the halos) of ordinary nearby galaxies (*5, 6, 7, 8*)**.** By the 1980's, the evidence for dark matter with an abundance of about 20% of the total energy density was widely accepted, although the nature of the dark matter remained a mystery.

[1]University of Cambridge. [2]Princeton University.

After the introduction of inflationary theory or the very early universe by Guth (*9*), many theoretical cosmologists became convinced that the universe must be flat and that the total energy density must equal the value (termed the critical value) that distinguishes a positively curved, closed universe from a negatively curved, open universe. Furthermore, noting how the evidence for dark matter was growing and extrapolating from the previous decade of study, the theoretical cosmologists became attracted to the beguiling simplicity of a universe in which virtually all of the energy density consists of some form of matter, roughly 4% being the ordinary matter and 96% the dark matter. In fact, observational studies were never compliant to this vision. Although there was a wide dispersion in total mass density estimates, there never developed any convincing evidence that there was sufficient matter to reach the critical value. The discrepancy between observation and the favored theoretical model became increasingly sharp.

Finally, dark energy came to the rescue (*10*). The only thing dark energy has in common with dark matter is that both components neither emit nor absorb light. In all other respects, they are different. Microphysically, they are composed of different constituents. Most significantly, dark matter, like ordinary matter, is gravitationally self-attractive and clusters with ordinary matter to form galaxies. Dark energy is gravitationally self-repulsive and remains nearly uniformly spread throughout the universe. Hence, a census of the energy contained in all the galaxies would miss almost all of the dark energy. So, by positing the existence of a dark energy component, it became possible to account for the 70-80% discrepancy between the measured mass density and the critical energy density predicted by inflation (*11, 12, 13, 14*).

But the dark energy dominated models make a strong prediction – that the universe is currently accelerating, due to the gravitational self-repulsion of the dominant dark energy component. This ran contrary to the then-current best observational tests based on the brightness of distant supernovae. Then, two independent groups (*15, 16*) found evidence of the acceleration from observations of supernovae, and the model with a dominant dark energy component became the concordance model of cosmology.

Dark energy has changed our view of the role of dark matter in the universe and our vocabulary for describing the cosmological possibilities. If this paper had been written a decade ago, before any serious consideration of dark energy, the focus would have been on the mass density. According to Einstein's general theory of relativity, in a universe composed only of matter (particles and radiation), it is the mass density that determines the geometry, the past history and the future evolution of the universe. For example, if the mass density exceeds the critical value, the self-gravity of the matter would cause the current expansion to eventually halt and reverse and, also, space would be positively curved. If the mass density is right at the critical value, space is flat (Euclidean) and the universe expands forever. Hence, the structure and fate of the universe would rest on the value of the ordinary plus dark matter density. With the addition of a new component, the story is totally different. First, what determines the geometry of the universe is whether the total *energy* density equals the critical value, where now we add to the mass contribution (identifying its energy according to $E=mc^2$) the dark energy contribution. Second, the period of matter domination has given way to dark energy domination. So, the important cosmological role of dark matter is in the past when it was the dominant contribution to the energy density, roughly the first few billion years. Our future is determined by the nature of the dark energy, which is sufficient to cause the current expansion of the universe to accelerate, and the acceleration will continue unless the dark energy should decay or change its equation of state.

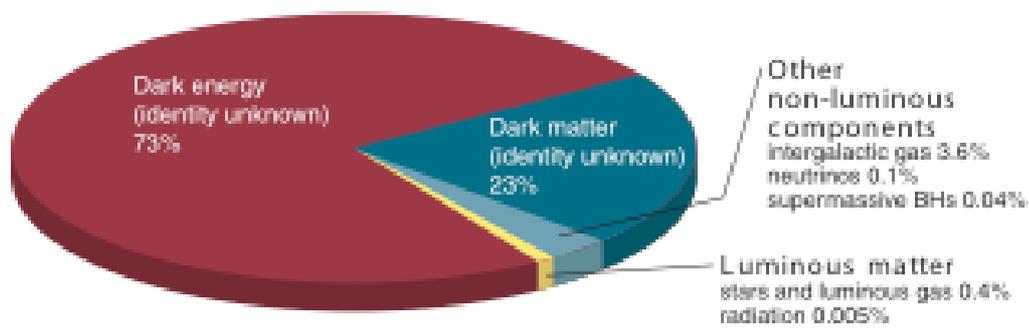

**Figure 1.** The luminous (light-emitting) components of the universe only comprise about 0.4% of the total energy. The remaining components are dark. Of those, roughly 3.6% are identified: cold gas and dust, neutrinos, and black holes. About 23% is dark matter, and the overwhelming majority is some type of gravitationally self-repulsive dark energy.

We have neglected one very important sub-plot up to this point – dark matter as the agent producing the growth of cosmic structure. We would not exist today were it not for the dark matter, which played a crucial role in the formation of the present structure in the universe. Without dark matter, the universe would have remained too uniform to form the galaxies, stars and planets. The universe, while nearly homogeneous and isotropic on its largest scales, shows a bewildering variety of structures on smaller scales: stars, galaxies, clusters of galaxies, voids and great walls of galaxies have been found. The only known force capable of moving matter on such large scales is Newton's gravity. And since, in a smooth and uniform medium, there will be no irregularities to produce gravitational forces, all structures must have been seeded by small fluctuations imprinted on the universe at very early times. These fluctuations should leave a signature on the CBR radiation left over from the big bang. Ordinary matter could not produce fluctuations to create any significant structures without leaving a signal bigger than what was observed in the CBR, because it remains tightly coupled to radiation, preventing it from clustering, until recent epochs.

On the other hand, dark matter, which is not coupled to photons, would permit tiny fluctuations (consistent with the CBR observations) to grow for a long, long time before the ordinary matter decoupled from radiation. Then, the ordinary matter would be rapidly drawn to the dense clumps of dark matter and form the observed structure. There would still need to be initial fluctuations, but their amplitude could be substantially smaller than otherwise. In 1991, the COBE satellite team announced the successful detection of these fluctuations, confirming an entirely independent argument for the existence of dark matter.

The required material was called Cold Dark Matter (CDM), since, in addition to its properties of invisibility and gravitational self-attraction, it was necessary for it to consist of non-relativistic particles to produce the observed structure and these, for simplicity, were assumed to contain no internal thermal motions, *i.e.,* they were cold.

A final important ingredient in the standard paradigm must be mentioned before we can begin to assess the validity of the picture. The initial spectrum of perturbations (ratio of long waves to short waves) must be specified in order to predict the gravitational effects of these waves. The inspired guess proposed independently by physicists Harrison, Peebles and Zeldovich in the 1970's was that the initial density fluctuations were "scale-invariant." That is, if we decompose the energy distribution into a sum of sinusoidal waves of varying wavelength, the wave amplitudes of the waves were the same for all wavelengths. One of the great triumphs of the inflationary scenario (*17, 18, 19, 20, 21*) is that it provided a well-motivated, dynamical mechanism for producing a nearly scale-invariant (defined by "spectral index": *n = 1*) spectrum. This prediction has now been confirmed by the WMAP satellite team, which found $n = 0.99 +- 0.04$ (*22*). Hence, it seems that our observational data has now confirmed our basic hypotheses and provided detailed information about the amount of dark matter in the universe and the initial distribution of matter and energy.

This provides a complete prescription for the conditions that lead to the growth of gravitational perturbations and the development of the universe to the state in which we now see it. It should be a matter of straightforward, if complex, computation to see if the developments predicted by this model agree with the universe as seen around us and also with its earlier phases as seen through the time machine made possible by powerful telescopes peering back into earlier epochs.

But we cannot claim to understand the evolution of structure in the universe, if we do not know the nature of the dark matter and how it fits within our models of fundamental physics. But, there is more at stake than that. The mass, stability, and interactions of dark matter with itself and with ordinary matter will all affect how dark matter participates in the formation of structure in the universe. Two kinds of dark matter are already known, neutrinos and black holes, (*23*) but they are generally thought to be minor contributions. Although there are current favored candidates for the majority component, it is the precise nature of the dark matter that is currently the most uncertain and interesting issue. Here we explore these issues: the possible candidates, their implications for structure formation, and how we might use a combination of particle detectors and astronomical observations to resolve the nature of dark matter.

# THE FAVORED CANDIDATES FOR DARK MATTER

For over a decade, the favored candidates for dark matter have been hypothetical elementary particles that are *long-lived, cold and collisionless*. *Long-lived* means the lifetime must be comparable to or greater than the present age of the universe, about 14 billion years. *Cold* means that the particles are non-relativistic at the onset of the matter-dominated epoch so that they are immediately able to cluster gravitationally. Because clustering occurs on length scales smaller than the Hubble horizon (age of the universe multiplied by the speed of light) and the Hubble horizon was much smaller at matter domination than today, the first objects to form – clumps or halos of dark matter – were much tinier than the Milky Way and much less massive. As the universe expanded and the Hubble horizon grew, many of these first small halos merged to form larger scale structures, which later merged themselves to form yet larger scale structures. The result is a hierarchy of structure ranging over many orders of magnitude in volume and mass, in accordance qualitatively with what is observed. In contrast, *hot* relativistic particles, such as light, massive neutrinos, would be moving too fast at matter domination to gravitationally cluster, and would result in a dramatically different distribution of structure inconsistent with what is observed. Hence, it has been known for nearly 20 years that light neutrinos must be a negligible component of the dark matter mass density, a conclusion that has been recently supported by measurements of the neutrino mass in underground solar neutrino experiments. *Collisionless* means that the interaction cross-section between dark matter particles (and between dark matter and ordinary matter) is so small as to be negligible for densities found in dark matter halos. The particles are only gravitationally bound to one another and travel unimpeded in orbits in the halos with a broad spectrum of eccentricities.

Cold, collionless dark matter has been favored for several reasons. First, numerical simulations of structure formation with cold, collisionless dark matter agree with most observations of structure. Second, for a special subclass known as WIMPs (weakly interacting massive particles), there is a natural explanation for why they have the requisite abundance. If particles interact through the weak force, then they are in thermal equilibrium in the first trillionths of a second after the big bang, when the density and temperature are high, and, then, they fall out of equilibrium with a concentration that is simply predicted from their annihilation cross-section. For a weak force cross-section, the expected mass density today spans a range that includes 20-30% of the total energy density of the universe, as observed. A third reason for favoring cold, collisionless dark matter is that there are specific appealing candidates for the dark matter particles in models of fundamental physics.

One candidate is the *neutralino*, a particle that arises in models with supersymmetry. Supersymmetry, a fundamental aspect of supergravity and superstring theories, requires a (yet unobserved) boson partner particle for every known fermion and a fermion partner particle for every known boson. If supersymmetry were extant today, the partners would have the same mass. But, because supersymmetry is spontaneously broken at high temperatures in the early universe, today the masses are different. Also, most supersymmetric partners are unstable and have decayed soon after the symmetry breaking. However, there is a lightest partner (with mass of order 100 GeV) that is prevented by its symmetries from decaying. In the simplest models, these particles are electrically neutral and weakly interacting – ideal candidates for WIMPs. If the dark matter consists of neutralinos, then large, sensitive underground detectors can detect their passage through the Earth as our planet travels around the Sun and through the dark matter in our own solar neighborhood. There are numerous efforts underway today that are beginning to explore the likely range of mass and cross-section of neutralinos. However, that detection does not necessarily mean that the dark matter consists primarily of WIMPs, which might quite possibly be, like neutrinos, only a small subcomponent of the dark matter.

Another appealing candidate is the axion, a very light neutral particle (with mass of order 1 μeV) important in suppressing strong CP violation in unified theories. The axion interacts through such a tiny force that it is never in thermal equilibrium, so the explanation for its abundance is not as simple. It immediately forms a cold Bose condensate that permeates the universe. For the axion, also, detectors have been built and have been running for several years.

The facts that cold, collisionless dark matter is simple to parameterize, results in numerical simulations which agree with most observations, and is motivated by particle physics explain why it is the leading candidate. But, the real test is just beginning, as vast improvements in numerical simulation and observations in recent years are leading to much more precise tests.

CRACKS IN THE FOUNDATION

Because the concordance model, combined with the assumption of cold, collisionless dark matter, is mathematically quite specific (even if some of the parameters that enter into it are known imprecisely), it can be tested at many different physical scales. The largest scales (thousands of megaparsecs (Mpc) – one parsec is 3.26 light-years, a kiloparsec (kpc) is one thousand parsecs and an Mpc is one million parsecs) are seen in the CBR itself. These measure the primordial distribution of energy and matter when their distribution was nearly uniform and there was no structure. Next come measurements of the large-scale structure seen in the distribution of galaxies ranging from several Mpc to nearly one thousand Mpc. Typically, these measurements span concentrations of dark matter ranging from small to intermediate. Over all of these scales, observation and theory are consistent inspiring great confidence in the overall picture.

However, on smaller scales, from one Mpc down to the scale of galaxies, kpc, and below, there is confusion. Either the results of the tests are uncertain, or they indicate disagreement with the naïve expectations of the theory. These apparent disagreements began to surface several years ago (*24, 25, 26*) and no consensus has emerged as to whether or not they represent real problems. For the most part, theorists believe that, if there is a problem, it is much more likely to be due to our specific assumption about the nature of dark matter rather than a problem with the global picture given by the concordance model. That there should be more uncertainty about smaller objects that are relatively closer may seem puzzling at first. First, on large scales gravity is king, so an understanding of the predictions involves only very straightforward computations based on Newton's and Einstein's laws of gravity. On smaller scales, the complex hydrodynamical interactions of hot, dense matter must be included. Second, the fluctuations on large scales are very small (a few percent or less), and we have very accurate methods of computing such quantities. But, on the scales of galaxies, the physical interactions of ordinary matter and radiation are quite complex. Supercomputer simulations are required, but they are not yet entirely reliable or reproducible from one investigator to another. As the problems have emerged, there have been changes both in the observational domain and the claimed theoretical predictions, further complicating the situation. The principle purported problems found on smaller scales are as follows:

1) Substructure, small halos and galaxies orbiting within larger units, may not be as common as is expected on the basis of numerical simulations of cold collisionless dark matter:

    a) The number of halos expected varies roughly as the inverse of the mass, so many dwarf systems, similar to our companions, the Magellenic Clouds, should be found – far more than.
    b) The lensing effect of small halos should be evident from the distribution of brightnesses of multiple images of a given galaxy, but the current evidence is inconclusive (*27*).
    c) The small halos, spiraling into the Milky Way and other systems should puff up the thin discs of normal galaxies by more than is observed (*28, 29*)

2) The density profile of dark matter halos should exhibit a "cuspy" core in which the density rises sharply as the distance from the center decreases, in contrast to the central regions of many observed self-gravitating systems:

    a) Clusters of galaxies, as observed in studies of gravitational lensing, have less cuspy cores than computed models of massive dark matter halos (*30*).
    b) Ordinary spiral galaxies, such as our own, have much less dark matter in their inner parts than expected (*31, 32*), as do some low surface brightness systems (*33*).
    c) Dwarf galaxies, like our companion systems, Sculptor and Draco have nearly uniform density cores in contrast to the expected cuspy density profile (*34, 35*).
    d) Hydrodynamic simulations produce galaxy disks that are too small and have too little angular momentum compared to observations (*36*).
    e) Many high surface brightness spiral galaxies exhibit rotating bars, which are normally stable only if the core density is lower than predicted (*37)*.

Addressing these issues is complex, and we are still mired in uncertainty. With regard to (1a) it seems likely that the explanation for the relatively small number of faint galaxies lies in the physics of galaxy formation. The explanation (*38*), is that halos having a central potential less than or comparable to the ionization energy of hydrogen will not be able to retain photo-ionized gas, and form stars. Hence, they are effectively invisible and would not be counted by observers (*39, 40*). Current strong lensing estimates using the distribution brightness ratios of multiple images of a given galaxy to determine the amounts of small-scale structure orbiting a galaxy, say, (1b) are difficult to understand. On the one hand, seeming evidence for small halos been found; on the other hand, it suggests more small halos than expected. A reasonable conclusion is that the observed distribution of brightness ratios may be due to effects other than small halos. Item (1c) is even less easy to pin down; if the discs form late enough and the matter density is small enough, so that most infall occurs early, then the disruption and thickening of late forming galactic disks by infalling satellites is unimportant. In sum, the evidence with respect to amounts of substructure observed *vs.* expected cannot be used at this time to argue either for or against LCDM with much conviction.

The second set of objections, based on the cusp density expected for the inner parts of the cold collisionless dark matter also, is observationally somewhat stronger (2a) - (2e). There are definitely many systems that do not show the steep profiles or the high mass concentrations in the inner core, as noted above. Two notions have been raised for resolving the apparent discrepancies. First, there may be dynamical processes that occur through the interaction between dark matter and the baryonic matter near the core that could reduce the central dark matter concentrations (*41, 42*). These proposals, while ingenious, seem strained, and the physical mechanisms invoked would also tend to disperse the old and dense bulge or spheroid component in a fashion inconsistent with observations. Alternatively, maybe the theoretical predictions of a cuspy profile are not as certain as had been supposed (*43, 44, 45*)**.** There may be no discrepancy at least for smaller mass halos because cold, collisionless dark matter does not really lead cuspy inner cores after all for such systems. Better dark matter simulations must be performed before we can be sure about whether (2a)-(2c) are serious problems or not. On the other hand, the large angular momentum of galactic disks and the preponderance of barred galaxies is hard to explain. Overall, however, the evidence to date, taken in its totality, does indicate that there is a discrepancy between the predicted high densities and the observed much lower densities in the inner parts of dark matter halos, ranging from those in giant clusters of galaxies ($M \geq 10^{15}$ solar masses) to those in the smallest dwarf systems observed ($M \leq 10^9$ solar masses).

## ALTERNATIVES TO COLD, COLLISIONLESS DARK MATTER

The possible discrepancies between theory and observation have motivated new proposals for the nature of dark matter. Each proposed variation from standard cold, collisionless dark matter (**CCDM**) has two properties: (1) it can "solve" some or all of the problems described in the previous section, and (2) it leads to additional predictions that would distinguish it from all the other alternatives (see Section V). A non-exhaustive list of examples follows:

1. *Strongly Self-Interacting dark matter* (**SIDM**): The dark matter might have a significant self-scattering cross-section $\sigma$, comparable to the nucleon-nucleon scattering cross-section (*46*). Then, in any halo, large or small, where the number of particles per unit area (the surface density) $\times \sigma$ is greater than unity, collisions amongst the dark matter particles leads to a complex evolution of the structure. During the initial phases of this process, which lasts longer than the present age of the universe, the central densities decline in the desired fashion due to the scattering of dark matter particles. Also, scattering strips the halos from small clumps of dark matter orbiting larger structures, making them vulnerable to tidal stripping and reducing their number.

2. *Warm dark matter* (**WDM**): Dark matter may be born with a small velocity dispersion (*e.g.,* through decay of another species) (*47, 48*)**,** which leaves it now with only perhaps 100 m/s velocity but which can have a significant effect on small scale structure. Extrapolating back in time, this velocity increases to a value sufficient to have a significant effect on small-scale structure (since the particles are moving too fast to cluster gravitationally on these scales). There are fewer low mass halos and all halos have a less steep profile in the innermost core. Also, because most of the lowest mass halos are born by the fragmentation of larger structures in this picture, they are found in high density regions and the voids tend to be emptier of small systems than in the standard cold, collisionless dark matter scenario.

3. *Repulsive dark matter* (**RDM**): Dark matter may consist of a condensate of massive bosons with a short range repulsive potential (*49*). The inner parts of dark matter halos would behave like a superfluid and be less cuspy.

4. *Fuzzy dark matter* (**FDM**): Dark matter could take the form of ultra-light scalar particles whose Compton wavelength (effective size) is the size of galaxy core (*50*)**.** Therefore, the dark matter cannot be concentrated on smaller scales, resulting is softer cores and reduce small-scale structure.

5. *Self-Annihilating dark matter* (**SADM**): Dark matter particles in dense regions may collide and annihilate, liberating radiation (*51*)**.** This reduces the density in the central regions of clusters for two reasons: direct removal of particles from the center and re-expansion of the remainder as the cluster adjusts to the reduced central gravity.

6. *Decaying dark matter* (**DDM**): If early dense halos decay into relativistic particles and lower mass remnants, then core densities, which form early, are significantly reduced without altering large scale structure (*52*).

7. *Massive Black Holes* (**BH**): If the bulk of the dark matter in galactic halos were in the form of massive black holes with mass of about one million solar masses, then several dynamical mysteries concerning the properties of our galaxy could be better understood (*53*). In normal galaxies dynamical friction between the massive black holes and the ordinary matter would cause those in the central few kiloparsecs to spiral into the center, depleting those regions of dark matter and providing the ubiquitous central massive black holes seen in normal galaxies.

While all of these ingenious suggestions were designed to reduce the central densities of dark matter halos, they achieve this end in different ways, and they should have different observational signatures. This provides ways of classifying the alternatives and devising tests that would enable us to eliminate some of the alternatives and further constrain the remaining ones.

# DETERMINING THE NATURE OF DARK MATTER

At first sight, the conceivable alternatives to cold collisionless dark matter are so numerous that it may seem impossible ever to distinguish among them However, the story turns out to be a happy one in that each alternative produces distinctive modifications on small scales that can be tested through improved astronomical observations and numerical simulations. The local universe – the small objects that orbit galaxies and the galaxy cores – turns out to be a marvelous new laboratory for examining the nature of dark matter.

The predictions of the various alternatives are distinctive because their modifications to the cold collisionless picture depend on different physical properties. **SIDM**, **BH** or **SADM** only affect halos when the *interaction rate* rises above a certain threshold value. The interaction rate depends on the *surface density* if the cross-section is velocity-independent or, more generally, the product of the *cross-section* and *velocity*. In all these cases, the interaction effect is slow because (by design) only a few scatterings take place within the lifetime of the Universe. **WDM**, **RDM**, or **FDM** all proposals that have a built-in characteristic *length scale* below which dark matter halos are affected. **DDM** has a characteristic built-in time *scale* after which dark matter halos are affected on all length scales and for all surface densities.

The alternatives also alter the history of structure formation compared to the cold collisionless dark matter picture in different ways. **SIDM** maintains the same sequence of structure formation but slowly rearranges the distribution of dark matter in dense regions. **SADM** is similar, except that it removes dark matter altogether from dense regions. Depending on details, **RDM** and fuzzy **FDM** may or may not affect the sequence of structure formation, either, but they insure that the smaller scale objects are forced to have a low physical density. **DDM** removes dark matter on all scales beginning after a characteristic decay time; because a lot of mass is lost through the decays, a higher rate of clustering is required throughout to match the observed galaxy cluster masses and match the other proposals. **WDM** delays the onset of structure formation until the dark matter cools sufficiently to gravitationally cluster, initially suppressing small scale structure formation but then creating it later by the fragmentation of larger scale structures. Finally, the **BH** alternative requires that there be significant non-linear structure on one million solar mass scales built-in *ab initio*, rather than grown from small fluctuations.

Because of these differences, the candidates for dark matter each face distinctive constraints and challenges. If the cross-section is too large, self-interaction (or self-annihilation) could lead to the evaporation of the halos of galaxies in clusters, in conflict with observation (*32, 54*). For **WDM**, for which structure formation is delayed compared to the standard picture, evidence for early galaxy and star formation provides a strong constraint. If the high electron-scattering optical depth apparently found by WMAP is confirmed,(an indicator of significant star formation at very early epochs), there would not be room for any delay (*22, 55*). Similarly, **SADM** could potentially destroy all small halos made at early times before they become sites for new small galaxies. A challenge for **DDM** is that it seems to require a higher production of massive, dense clusters in the early universe than observed in order to obtain the right mass distribution after decay.

There may be new kinds of observations that can distinguish among the candidates for dark matter by taking advantage of their qualitative differences, as we discuss below. To be quantitative in our predictions, detailed numerical simulations of each case are necessary and we would urge that these be done in the near future. We would not be surprised if some of the guesses we are putting forward will turn out to be incorrect when accurate calculations are made.

First we consider the epoch at which objects of different mass will form in the different scenarios (*Fig 1*). To give the same structures today, objects of a given mass will need to form earlier in the **DDM**, **SADM**, and **BH** scenarios as compared to the standard **CCDM** and **SIDM** scenarios. The low mass objects will form later in at least some **FDM** and **RDM** scenarios, and, in the **WDM** scenario, they will form later and only by fragmentation of more massive objects. The mass of, and even the existence of low mass galaxies at early times will provide a valuable diagnostic to distinguish the alternatives: the WMAP observations favors models which form structure at early times.

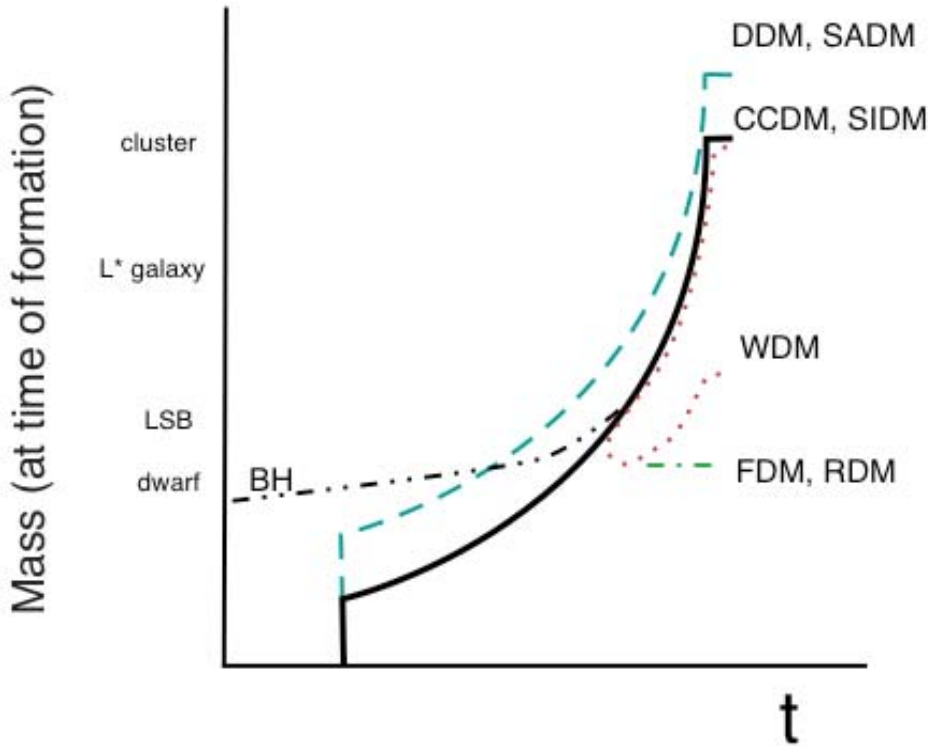

**Figure 2.** History of structure formation: the time of formation for objects of a given mass M (as measured at formation) for structures with increasing mass (dwarf, low surface brightness (LSB), ordinary (L*) galaxies and galaxy clusters) for different models of dark matter. Structure formation begins shortly after the onset of the matter dominated epoch (left hand side). Acronyms are explained in text.

Next we look at the demography expected to be seen in the local universe when population studies are made. How many small and how many large dark matter halos show exist is presented in *Fig. 2*. In the **WDM**, **FDM**, and **RDM** scenarios, small mass objects are underabundant compared to the **CCDM**, **SIDM**, and **SADM** scenarios and in the **BH** scenario, they are probably overabundant. **WDM** calculations (48) reveal that objects made by fragmentation are present but at a lower level. The small halos may be difficult to observe directly because they may be unable to retain gas long enough to make observable galaxies. But these small dark halos may be detected through their gravitational effects, such as lensing, puffing up of disks, and other dynamical interactions.

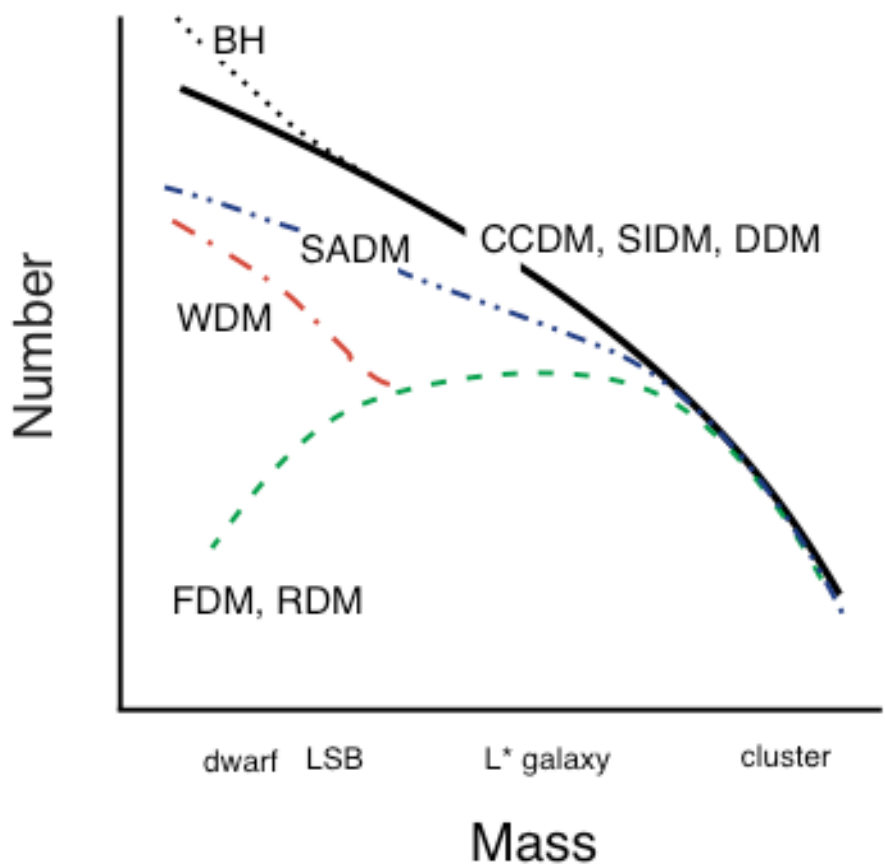

**Figure 3.** Demography: how the number of objects of a given type depends on their mass (as observed today) for different dark matter models.

The internal structure of the halos provides another feature to distinguish one model from another. In the **CCDM** model, low mass halos were made early when the universe was denser than later, and so they are themselves more dense than structures formed later. This is shown in their internal structure. So, *Figure 3* reflects the historical conditions shown in *Figure 1* but allows one to study nearby objects. This is a critical issue because the inner parts of dark matter halos do seem to be considerably less dense than expected in the standard **CCDM** model. Here the **BH** scenario is complex. For isolated dark matter halos, which do not contain baryonic components, the dynamical evolution will be qualitatively similar to that of star clusters. On a time scale proportional to the dynamical (or orbital) time multiplied by the ratio of the system mass to the typical black hole mass the inner profile will first flatten and then collapse via a process called the gravo-thermal instability. For parameters appropriate to galactic dark matter halos, even the first process will only occur for the lowest mass dwarf systems and thus less cuspy cores would be expected in the local dwarf galaxies. In normal galaxies the stronger interaction is between the black holes and the normal stellar component, and this leads, as noted before, to clearing out the black holes from the inner parts of the galaxies with them sinking to the center where they either merge or are ejected.

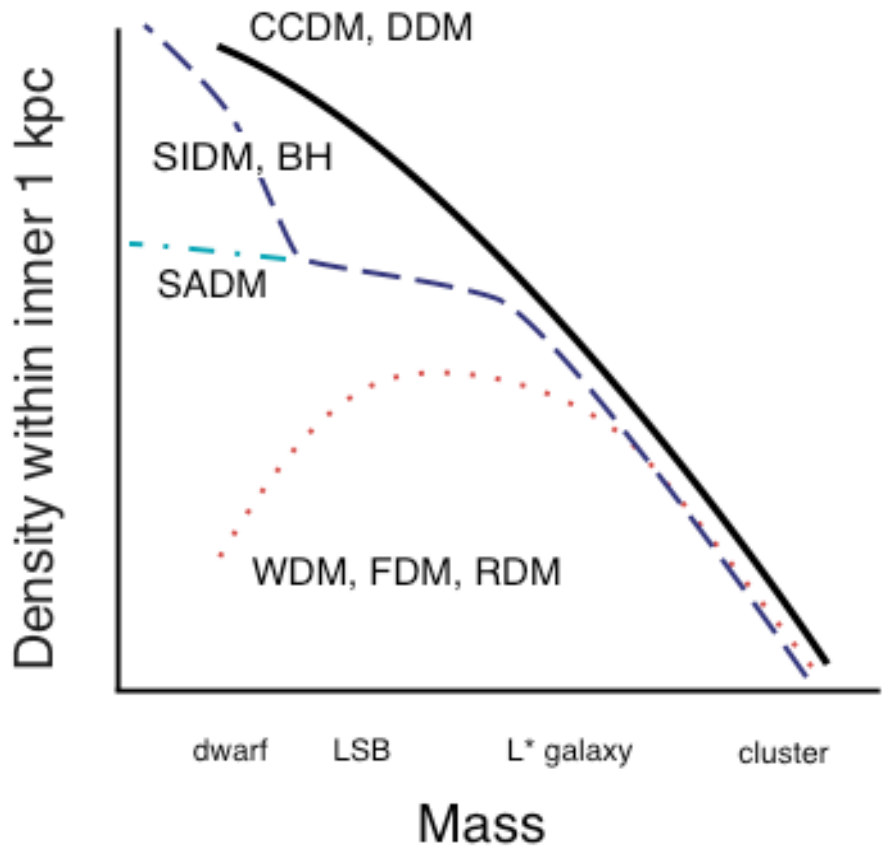

**Figure 4.** Internal structure: how the density density of the inner one kiloparsec depends on the mass of the system for different dark matter models.

Finally, we examine the environments within which different kinds of objects should be found. In the standard model, low mass halos will be distributed relatively more uniformly than the higher mass halos, so that the large voids seen in the distribution of massive galaxies should be populated with halos of low mass and perhaps also with associated low mass galaxies. To date, studies have not found such galaxies, but we do not yet know if this because of an absence of the predicted low mass halos in the voids or simply because the ones that are there have not been able to make galaxies. In the **WDM** scenario, the low mass halos are typically near the high mass ones as they form by fragmentation of larger structures. For the **SIDM**, **SADM**, **FDM** and **RDM** scenarios, the abundance of low mass objects will decline in the vicinity of the highest mass ones. In **SIDM**, it will be because interactions will boil away the cooler low mass halos by direct particle-particle collisions, and, in the other three cases, it is because the low mass halos will have a low internal density and be fragile, hence easily shredded in tidal encounters with their bigger brothers. For the **BH** scenario, the voids would be heavily populated with small dark matter systems, but these might or might not contain observable stellar systems.

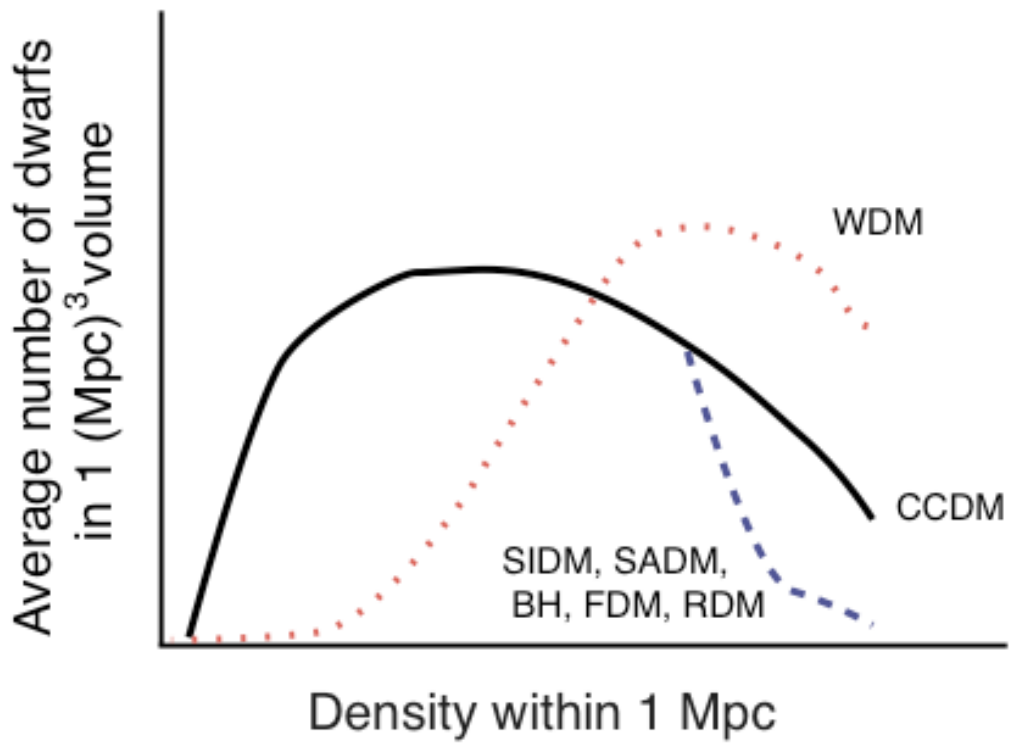

**Figure 5.** Environment: how the number of dwarfs in $(1 \text{ Mpc})^3$ volume depends on the average density within that volume.

CONCLUSIONS

The idea, that some mysterious "dark matter" dominates over the ordinary chemical elements, first broached by Fritz Zwicky over 65 years ago, is now the common wisdom, confirmed by many different lines of evidence. For most astronomical observations the simplest possible choice seems to give an adequate description: the dark matter is primarily made up of elementary particles which are *long-lived, cold and collisionless* and has been termed cold dark matter. The most direct way to see if this choice is correct is via earth based laboratory particle detectors and several experiments are underway.

But there are a variety of clues telling us that the world may not be as simple as the **CCDM** model. While the **CCDM** model is able to correctly predict observations made from the largest cosmological scales down to roughly those of galactic scale and from the early universe to the present epoch, there are many indications that on sub-galactic scales it predicts that there should be more dark matter than is detected gravitationally. Numerical simulations seem to predict that all galaxies should contain cuspy cores, where the density of dark matter rises sharply with decreasing radius, and most observations do not confirm this prediction. We need more accurate simulations and more accurate observations to see if these apparent discrepancies are real. If they are, then there are several interesting suggestions which could account for the less cuspy cores and, more importantly, would lead to predictions of other observables that could be used to test the variant pictures. These include the history of dark halo formation, the demography (mass distribution) of low mass halos, the detailed interior density distribution of galaxy halos and the environments within which different kinds of astronomical objects are found. We have sketched out the kinds of astronomical tests that could be made to narrow the search, but if history teaches us anything it is that the next important clues will come from a surprising direction. For example, it may be that our assumption of a single dominant component is simplistic. Some observation or calculation will be made that will reorient our inquiries and, if it happens as has happened so often in the past, we will realize that the important evidence has been sitting unnoticed under our noses for decades.

# REFERENCES

1. L. Wang, R. R.Caldwell, J. P. Ostriker, P. J. Steinhardt, *ApJ*. **530**, 17 (2000).

2. C. L. Bennett *et al., Astro-ph*/0302207 (2003).

3. K. A. Olive, G. Steigman, T. P. Walker, *Physics Reports* **333**, 389 (2000).

4. F. Zwicky, *ApJ* **86**, 217 (1937).

5. M. S. Roberts, A. H. Rots,. *Astr. and Astrophs.* **26**, 483 (1973).

6. J. P. Ostriker, P.J.E. Peebles, A. Yahil, *ApJ Letters* **193**, L1 (1974).

7. J. Einasto, A. Kaasik and E. Saar, *Nature* **250,** 309 (1974).

8. V. C. Rubin, N. Thonnard, W. K. Ford Jr., *ApJ. Letters* **225**, L107 (1978).

9. A. H. Guth, *Phys. Rev. D***23**, 347, (1981).

10. R. Kirshner, *Science* **300**, 1914 (2003).

11. P. J. E. Peebles, *ApJ* **284,** 439 (1984).

12. G. Efstathiou, W.J. Sutherland, S.J. Maddox, *Nature* **348**, 705 (1990).

13. L. Krauss, M. S. Turner, *Gen. Relativ. Gravit.* **27**, 1137 (1995).

14. J. P. Ostriker, P. Steinhardt, *Nature* **377**, 600 (1995).

15. A. Reiss, A. *et al.*, *Astron. J.* **116**, 109 (1998).

16. S. Perlmutter *et al.*, *ApJ*. **517**, 565 (1999).

17. J. Bardeen, P. J. Steinhardt and M. S. Turner, *Phys. Rev. D* **28**, 679 (1983).

18. A. H. Guth and S.-Y. Pi, *Phys. Rev. Lett.* **49**, 1110 (1982)

19. S. W. Hawking, *Phys. Lett. B***115**, 295 (1982).

20. V. F. Mukhanov, G. V. Chibisov, *JETP Letters*, **33**, 532 (1981).

21. A. A. Starobinskii, *Phys. Lett. B***117**, 175 (1982)

22. D. N. Spergel *et al.*, *Astro-ph/*0302209 (2003).

23. M. Begelman, *Science* **300**, 1898 (2003).

24. J. F. Navarro, C. S. Frenk, S. D. M. White, *ApJ* **490,** 493 (1997).

25. B. Moore, F. Governato, T. Quinn, J. Stadel, G. Lake, *ApJ Letters* **499,** L5 (1998).

26. A. V. Kravtsov, A. A. Klypin, J.S. Bullock, *ApJ* **502,** 48 (1990).

27. N. Dalal, C. S. Kochanek, *ApJ* **572**, 25 (2002).

28. G. Toth, J. P. Ostriker, *ApJ*. **389**, 5 (1992).

29. A. S. Font, J. F. Navarro, J. Stadel, T. Quinn, *ApJ* **563**, L1 (2001).

30. J. A. Tyson, G. P. Kochanski, I. P. Dell'Antonio, *ApJ. Lett.* **498**, L107 (1998).

31. J. J. Binney, N. W. Evans, *MNRAS*, **327**, L27 (2001).

32. R. Davé, D. N. Spergel, P. Steinhardt, B. Wandelt, *ApJ* **547**, 574 (2001).

33. F. C. Van den Bosch, B. E. Bobertson, J. J. Dalcanton, W.J.G. de Bok, *A.J* **119**, 1579 (2000).

34. F. Stoehr, S.D.M. White, G. Tormen, V. Springel, *MNRAS* **335**, L84 (2002).

35. J. T. Kleyna, M. Wilkinson, G. Gilmore, W. N. Evans, *Astro-ph/*0304093 (2003).

36. J. F. Navarro, M. Steinmetz, *ApJ* **528**, 607 (2000).

37. V. P. DeBattista, J. A. Sellwood, *ApJ* **493**, L5 (1998).

38. J. Silk, *ApJ*. **211**, 638 (1977).

39. W. A. Chiu, N. Y. Gnedin, J. P. Ostriker, *ApJ* **563**, L21 (2001).

40. K. Nagamine, M. Fukugita, R. Cen, J. P. Ostriker, *MNRAS* **327**, L10 (2001).

41. O. Y. Gnedin, H. S. Zhao, *MNRAS* **333**, 299 (2002).

42. M. D. Weinberg, N. Katz, *ApJ*. **580**, 627 (2002).

43. C. Power, J. F. Navarro, A. Jenkins, C. S. Frenk, S.D.M. White, V. Springel, J. Stadel, T. Quinn, *MNRAS* **338**, 14 (2003).

44. S. Ghigna, B. Moore, F. Governato, G. Lake, T. Quinn, J. Stadel, *ApJ* **544**, 616 (2000).

45. M. Ricotti, A*stro-ph*/0212146 (2002).

46. D. N. Spergel, P. J. Steinhardt, *Phys. Rev. Lett*. **84**, 3760 (2000).

47. P. Colín, V. Avila-Reese, O. Valenzuela, *ApJ* **542**, 622 (2000).

48. P. Bode, J. P. Ostriker, N. Turok, *ApJ* **556**, 93 (2001).

49. J. Goodman, *New Astronomy* **5**, 103 (2000).

50. W. Hu, R. Barkana, A. Gruzinov, *Phys. Rev. Lett*. **85**, 1158 (2000).

51. L. Kaplinghat, L. Knox, M. S. Turner, *Phys.Rev.Lett*. **85**, 3335 (2000).

52. R. Cen, *ApJ* **546**, L77 (2001).

53. C. Lacey, J.P. Ostriker, *ApJ* **229**, 633 (1985).

54. J. F. Hennawi, J. P. Ostriker, *ApJ* **572**, 41 (2002).

55. N. Yoshida, V. Springle, S.D.M. White, G. Tormen, *ApJ Lett.* **544**, L87-L90 (2000).